\documentstyle[aps,prd,multicol,epsf,epsfig]{revtex}\begin{document}

\newcommand{\be}{\begin{equation}}
\newcommand{\ee}{\end{equation}}
\newcommand{\ba}{\begin{eqnarray}}
\newcommand{\ea}{\end{eqnarray}}
\title{Centaurus A as the Source of ultra-high energy cosmic rays?}
\author{ Claudia Isola$^a$, Martin Lemoine$^b$, G{\"u}nter Sigl$^b$}
\address{$^a$ Centre de Physique Th{\'e}orique,\\
Ecole Polytechnique, 91128 Palaiseau Cedex, France\\
$^b$ Institut d'Astrophysique de Paris, C.N.R.S., 98 bis boulevard
Arago, F-75014 Paris, France}

\date{April 13, 2001}
\maketitle

\vspace{0.5truecm}
\begin{abstract}
We present numerical simulations for energy spectra and angular
distributions of nucleons above $10^{19}$~eV injected by the
radio-galaxy Centaurus~A at a distance 3.4~Mpc and propagating in
extra-galactic magnetic fields in the sub-micro Gauss range. We show
that field strengths $B\simeq0.3\mu{\rm G}$, as proposed by Farrar \&
Piran, cannot provide sufficient angular deflection to explain the
observational data. A magnetic field of intensity $B\simeq1\mu{\rm G}$
could reproduce the observed large-scale isotropy and could marginally
explain the observed energy spectrum. However, it would not readily
account for the $E=320\pm93\,$EeV Fly's Eye event that was detected at
an angle 136$^o$ away from Cen--A.  Such a strong magnetic field also
saturates observational upper limits from Faraday rotation
observations and X-ray bremsstrahlung emission from the ambient gas
(assuming equipartition of energy). This scenario may already be
tested by improving magnetic field limits with existing instruments.
We also show that high energy cosmic ray experiments now under
construction will be able to detect the level of anisotropy predicted
by this scenario.  We conclude that for magnetic fields
$B\simeq0.1-0.5\,\mu$G, considered as more reasonable for the local
Supercluster environment, in all likelihood at least a few sources
within $\simeq10\,$Mpc from the Earth should contribute to the observed
ultra high energy cosmic ray flux.
\end{abstract}
\vspace{1truecm}

\begin{multicols}{2}

\narrowtext

\section{Introduction}
In acceleration scenarios ultra high energy cosmic rays (UHECRs) with
energies above $10^{18}\,$eV are
assumed to be protons accelerated in powerful astrophysical sources.
During their propagation, for energies above $\gtrsim50$ EeV ($1EeV =
10^{18} eV$) they lose energy by pion production and pair production
(protons only) on the microwave background. For sources further away
than a few dozen Mpc this would predict a break in the cosmic ray flux
known as Greisen-Zatsepin-Kuzmin (GZK) cutoff~\cite{gzk}, around
$50\,$EeV. This break has not been observed by experiments such as
Fly's Eye\cite{Fly}, Haverah Park\cite{Haverah},
Yakutsk\cite{Yakutsk}, Hires\cite{Hires} and AGASA\cite{AGASA}, which
instead show an extension beyond the expected GZK cutoff and events
above $100\,$EeV.  This situation has in recent years triggered many
theoretical explanations ranging from conventional acceleration to new
physics as well as the construction of large new
detectors~\cite{reviews}.

In bottom-up scenarios of UHECR origin, in which protons are
accelerated in powerful astrophysical objects such as hot spots of
radio galaxies and active galactic nuclei~\cite{biermann},
one would expect to see the source in the direction of
arrival of UHECRs. The lack of observed counterparts to the highest
energy events~\cite{ssb,ES95} implies the existence of large scale intervening
magnetic fields with intensity $B\sim0.1-1\,\mu$G~\cite{ES95}, which
would provide sufficient angular deflection, or bursting sources and a
magnetic field of intensity $B\gtrsim10^{-11}\,$G which would impart
sufficient time delay to UHE protons to explain their lack of
correlation in time of arrival with optical or high energy
photons~\cite{WM-E96}. It has been realized recently that magnetic
fields as strong as $\simeq1\mu\,$G in sheets and filaments of large
scale structure, such as our Local Supercluster, are compatible with
existing upper limits on Faraday rotation~\cite{vallee,rkb,bbo}.  

Such strong magnetic fields would have profound consequences for the
propagation of charged ultra-high energy cosmic
rays~\cite{BO98,SLB99,LSB99}. In particular in the presence of a
magnetic field of strength $B\simeq0.1\,\mu$G, and for a source at
distance $\simeq10\,$Mpc, UHECRs with energy $E\lesssim100\,$EeV
would diffuse, while higher energy cosmic rays would propagate
rectilinearly. The resulting modification of the energy spectrum would
explain naturally the observed spectrum for a unique power-law
injection of index $\simeq2.2$. In Ref.\cite{LSB99}, we further showed
that a large number of such sources in the Local Supercluster, at
distance scale $\simeq10\,$Mpc and an ambient magnetic field
$B\simeq0.1\,\mu$G would explain the large scale isotropy of arrival
directions observed by AGASA\cite{AGASA}. It would also explain the
observed small-scale angular clustering (five doublets and one triplet
within $2.5^o$ out of 57 events above $40\,$EeV) by magnetic lensing
effects through the large scale turbulent magnetic field.

At first sight this suggests the possibility of having only one
object in the Sky as
the source of all UHECRs with $E\gtrsim5$EeV, including the
highest energy events. Two versions of such single source scenarios
have recently been put forward in the literature: one with an extreme
version of a coherent Galactic magnetic wind structure in which all
observed UHECRs supposedly can be traced back to M87 in the Virgo
cluster~\cite{ambs}, and a second one with Centaurus A at 3.4~Mpc
distance with an all-pervading magnetic field of intensity
$B\simeq0.3\,\mu$G~\cite{fp2}. In the present paper we examine
critically the latter of these scenarios using detailed numerical
simulations for the energy spectrum and the angular distribution of
UHECRs propagating in magnetic fields of r.m.s. strength
between 0.3 and 1 $\mu$G.

\section{Numerical Simulations}
Trajectories are calculated from the Lorentz equation in a given
magnetic field and pion production is treated as stochastic energy
loss while pair production is included into the equations of motion as
a continuous energy loss term.  This numerical tool has been used and
discussed in earlier publications~\cite{SLB99,LSB99}.

\begin{figure}
\centering\leavevmode \epsfxsize=3.5in \epsfbox{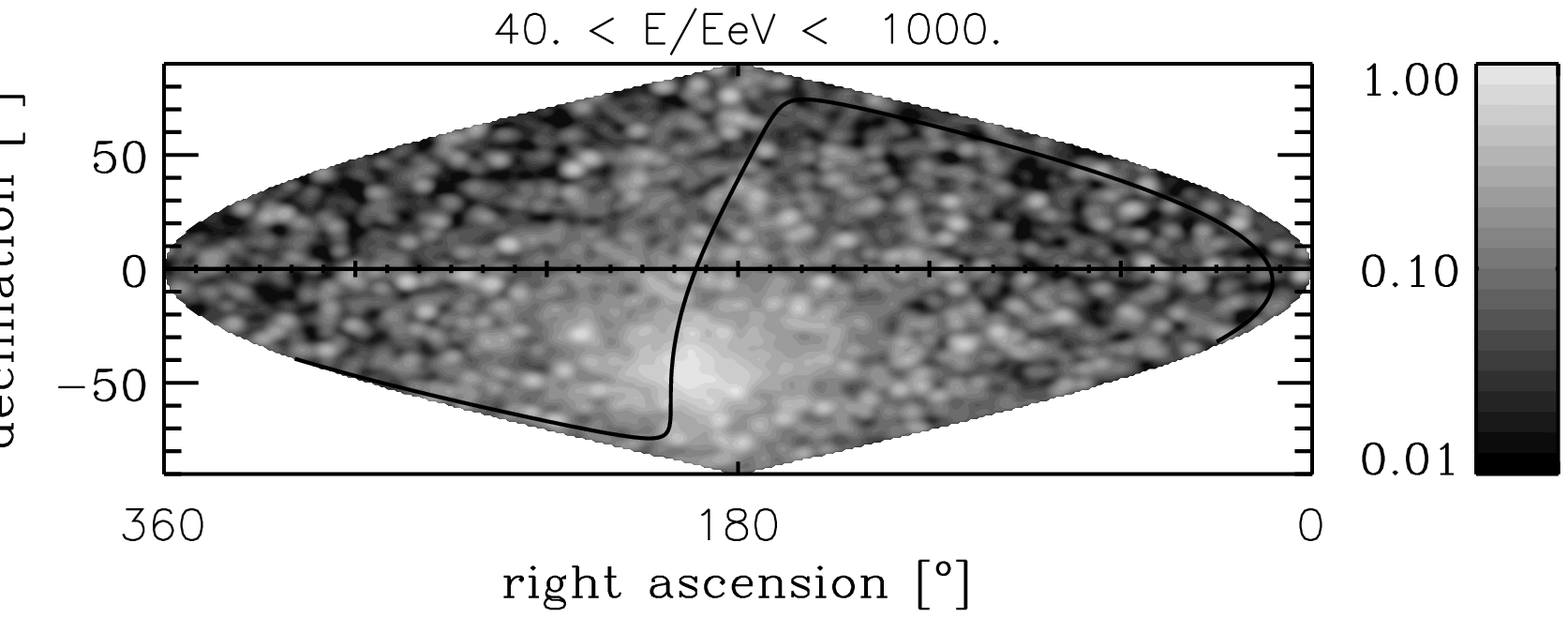}
\epsfxsize=3.5in \epsfbox{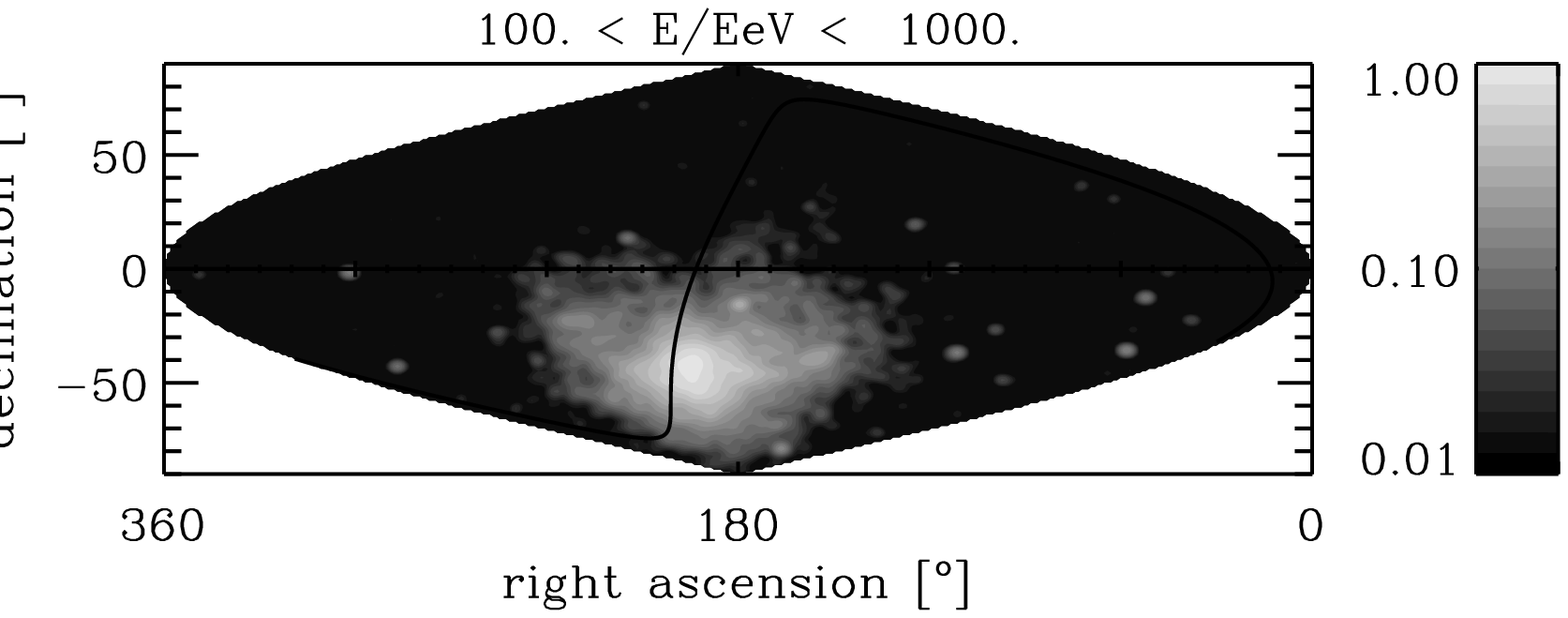}
\caption[...]{The angular image in terrestrial coordinates, averaged over
all 20 magnetic field realizations of 5000 trajectories each, for
events above $40\,$EeV (upper panel) and above $100\,$EeV (lower
panel), as seen by a detector covering all Earth, for the case
suggested in Ref.~\cite{fp2} 
corresponding to $B=0.3\,\mu$G, and the
source Cen--A located 3.4~Mpc away.  The grey scale represents the
integral flux per solid angle. The solid line marks the Supergalactic
plane.  The pixel size is $1^\circ$; the image has
been convolved to an angular resolution of 2.4$^\circ$ corresponding to
AGASA.}
\label{F1}
\end{figure}

We assume a homogeneous random turbulent magnetic field with power
spectrum $\langle B(k)^2\rangle\propto k^{n_B}$ for
$2\pi/L<k<2\pi/l_c$ and $\langle B^2(k)\rangle=0$ otherwise. We use
$n_B=-11/3$, corresponding to Kolmogorov turbulence, in which case
$L$, the largest eddy size, characterizes the coherence length of the
magnetic field; we use $L\simeq1\,$Mpc, which corresponds to a few
turn-arounds in a Hubble time. Physically one expects $l_c\ll L$, but
numerical resolution limits us to $l_c\gtrsim0.008L$. We generally use
$l_c\simeq0.03\,$Mpc, but we checked by increasing the resolution for
several runs that it has no effect on the results discussed in the
following. The magnetic field modes are dialed on a grid in momentum
space according to this spectrum with random phases and then Fourier
transformed onto the corresponding grid in location space. The
r.m.s. strength $B$ is given by
$B^2=\int_0^\infty\,dk\,k^2\left\langle B^2(k)\right\rangle$.

Typically, 5000 trajectories are computed for each magnetic field
realization obtained in this way for 10-20 realizations in total.
Each trajectory is followed for a maximal time of 10 Gyr and
as long as the distance from the observer is smaller than double
the source distance. We have also checked that the results do not
significantly depend on these cut-offs. Furthermore, the distance
limit is reasonable physically as it mimics a magnetic field
concentrated in the large scale structure, with much smaller
values in the voids, as generally expected.

\begin{figure} 
\centering\leavevmode 
\epsfxsize=3.5in
\epsfbox{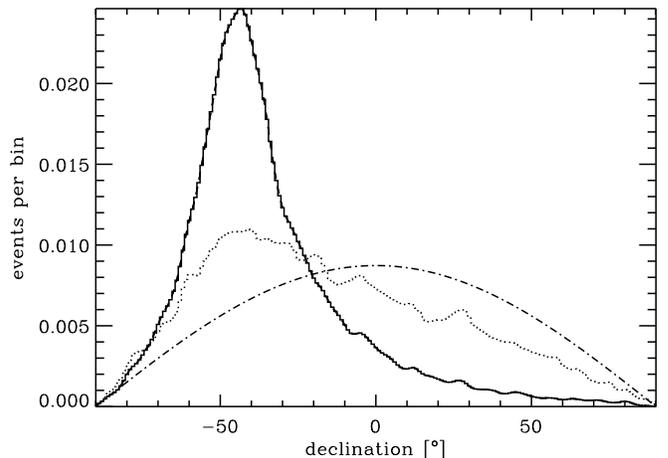} 
\caption{The distribution of arrival declination on Earth, averaged over
many realizations,  for $E\geq40\,$EeV (dotted line) and $E\geq100\,$EeV
(solid line), for the scenario corresponding to Figs.~\ref{F1}.
The dash-dotted line represents an isotropic
distribution. The pixel size is $1^\circ$ and the image has
again been convolved with an angular resolution of 2.4$^\circ$.}
\label{F2}
\end{figure}

In Fig.~\ref{F1}, we show the angular distribution of UHE events as
seen on Earth in equatorial coordinates, for two ranges of energies
$E\gtrsim40\,$EeV, and $E\gtrsim100\,$EeV, and for $B=0.3\,\mu$G, with
Cen--A as the source. These images are averaged over different spatial
realizations of the magnetic field. The distributions for specific
realisations are more anisotropic than the average due to cosmic
variance.  The angular distribution predicted by this one source model
is thus not consistent with the isotropic distribution deduced by
experimental data~\cite{Fly,AGASA}. This is even more clearly
demonstrated by Fig.~\ref{F2}, which gives the distribution in
declination of arrival coordinates of UHE events. The source Cen--A is
located at ${\rm RA}=201.3^o$, $\delta=-43.0^o$ in equatorial
coordinates, and corresponds to the peak of flux in
Figs.~\ref{F1},\ref{F2}.

\begin{figure}
\centering\leavevmode \epsfxsize=3.5in \epsfbox{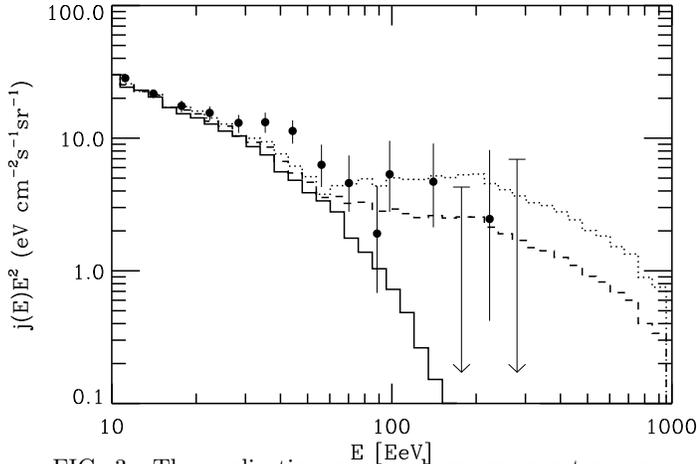}
\caption{The realization averaged energy spectra corresponding to
Figs.~\ref{F1},\ref{F2}. The solid line represents the spectrum that
would have been detected by AGASA, and has been obtained from
Eq.~(\ref{expos}). The dashed line indicates the spectrum uniformly
averaged over the whole sky.  The dotted line is the spectrum
predicted by an AGASA type experiment in the Southern hemisphere. The
one sigma error bars indicate the AGASA data. The various spectra have
been normalized to optimally fit the AGASA flux. Significant
uncertainties due to cosmic variance and parameters such as the
largest eddy size $L$ only occur for energies between $\simeq70\,$EeV
and a few hundred EeV where they are still smaller than a factor
$\simeq2.5$.}
\label{F3}
\end{figure}

Cen--A does not lie in the field of view of experiments that have
provided data so far such as the Fly's Eye and AGASA. This
and the fact that the angular deflection of particles with
$E\gtrsim100\,$EeV is relatively small has an important consequence
for the energy spectrum predicted by this model: The Northern hemisphere
experiments should never have detected the
highest energy events, for which the angular deflection is too weak to
bring the particle in the field of view. This is made clear in
Fig.~\ref{F3}, where we compare the spectrum predicted for
AGASA with the actually observed spectrum, assuming an injection
spectrum $\propto E^{-2}$ up to $10^{21}\,$eV. The prediction is obtained
by folding the simulated distribution in energy $E$ and arrival
direction $\Omega$, $D(E,\Omega)$, with the normalized
AGASA exposure function ${\rm AGASA}(\delta)$
(which to a good approximation only depends on declination $\delta$):
\begin{equation}
j(E)\equiv \int {\rm d}\, \Omega D(E,\delta){\rm AGASA}(\delta)
\label{expos}\,.
\end{equation}
For reference, we also show in Fig.~\ref{F3} the spectrum that would
be observed by an idealized detector covering the whole sky
uniformly, and by a mirror AGASA experiment located in the Southern
hemisphere. The solid angle integrated spectrum $\int{\rm d}\Omega
D(E,\delta)$ observed by a uniform detector is still different from the
injection spectrum ($\propto E^{-2}$) at low energies, while the two
match at high energies. This results from an increased local residence time
due to diffusion at low energies, and rectilinear propagation (hence
unaffected energy spectrum) at high energies~\cite{SLB99}. The pile-up
around $E\simeq40\,$EeV due to pion production of higher energy
particles, also contributes to the change of slope at low energies.

The scenario just discussed, which is already ruled out by the energy
spectrum and large-scale isotropy recorded by Northern hemisphere
detectors, has been proposed by Farrar \& Piran~\cite{fp2} on
analytical grounds to explain all observational data. The discrepancy,
as will be discussed in greater detail in Section~3, results from the
fact that they argued that diffusion held up to the highest energies,
whereas in fact the diffusion approximation breaks for
$E\gtrsim100\,$EeV, implying much larger anisotropies at these
energies. The impact of angular anisotropy on the energy spectrum for
a source located in the blind area of Northern hemisphere detectors
had also been overlooked in Ref.~\cite{fp2}.

For the scenario with $B=0.3\mu\,$G we have also established
a rough estimate of the minimal number of sources necessary to
explain existing observations in the following way: We overlayed
distributions from single sources in random directions (for
simplicity chosen to be at equal distances as Cen--A) and computed
the average and fluctuations of the declination distributions
of UHECR arrival directions. The minimal number of sources is
determined by requiring that the distribution is consistent
with isotropy within the fluctuations and turns out to be 5--10.

\begin{figure}
\centering\leavevmode 
\epsfxsize=3.5in
\epsfbox{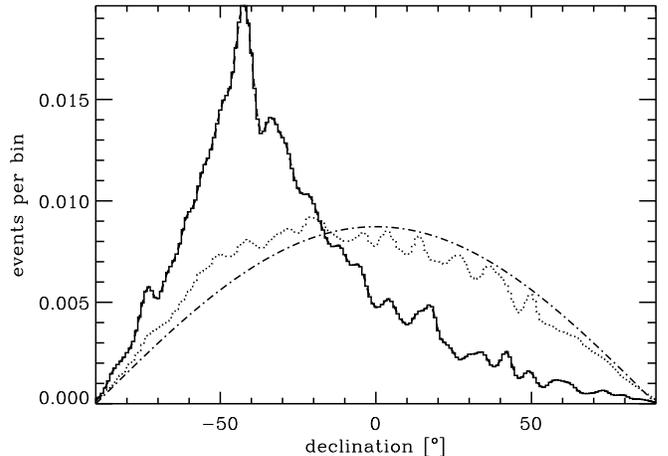}
\caption{The distribution of arrival declination on Earth,
averaged over many realizations, for $E\geq100\,$EeV (dotted line)
and $E\geq300\,$EeV (solid line). All other parameters are as in
Fig.~\ref{F2}, expect $B=1\,\mu$G.}
\label{F4}
\end{figure}

\begin{figure} 
\centering\leavevmode 
\epsfxsize=3.5in
\epsfbox{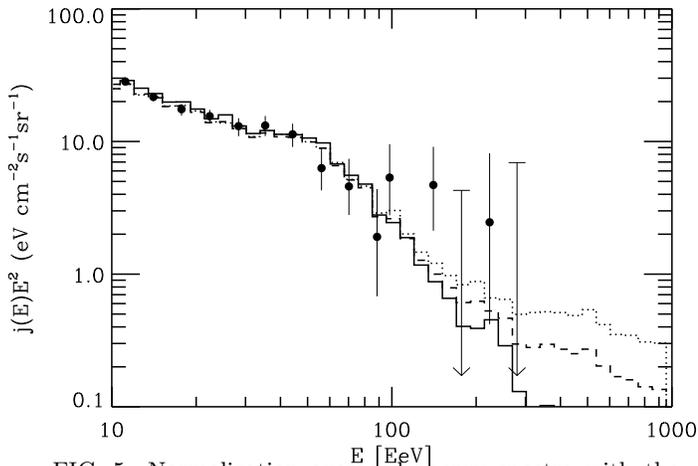}
\caption{Normalization averaged energy spectra with the same
conventions as in Fig.~\ref{F3}, but for $B=1\,\mu$G, all
other parameters being equal.}
\label{F5}
\end{figure}

We now investigate whether stronger magnetic fields, by providing
larger angular deflection, might provide a better match to the
observational data. In particular, we focus on the case where
$B=1\,\mu$G, and Cen--A is again the unique source of UHECRs.
The resulting arrival distribution in declination is shown for
several ranges of energies in Fig.~\ref{F4}, and the resulting
energy spectra, calculated as before, is shown in Fig.~\ref{F5}.
Increasing the magnetic field
strength increases the maximal energy at which diffusion takes place,
hence it decreases the anisotropy at each energy up to that maximal
energy, and thus reduces the differences between the spectra seen by
different detectors (AGASA, uniform, and Southern hemisphere
analogue of AGASA).

The large-scale anisotropy in this case is much smaller and could not
have been detected by Northern hemisphere experiments such as the
Fly's Eye and AGASA. The predicted energy spectrum for AGASA does
provide a very good fit to the observed spectrum, see Fig.~\ref{F5},
but the difference is not sufficient to rule out this scenario on this
ground. However, we note that it is unlikely that the highest energy
Fly's Eye event would have been detected in this model. This event of
energy $E=320\pm93\,$EeV makes an angle of $136^o$ with the line of
sight to Cen--A~\cite{ES95}. By folding the energy probability
distribution for the Fly's Eye event with the simulated distribution
of deflection angles and energies one finds that 95\% of events with a
recorded energy corresponding to the Fly's Eye event are deflected
less than $\simeq130^o$. 

Thus the present observational evidence is not sufficient to rule out
with a high degree of confidence the possibility that Cen--A is the
source of all ultra-high energy cosmic rays, {\it if} the intervening
magnetic field $B\gtrsim1\,\mu$G. Notably, one cannot rule out the
possibility of having a magnetic configuration different from
Kolmogorov turbulence, in which case the modification of the energy
spectrum would be different than what is shown in the present
simulations. For instance, it has been proposed that the magnetic
field is not all pervading, but strongly enhanced in regions close to
radio-galaxies that were active in the past~\cite{ES95,M-TE00}, in
which case the configuration seen by UHECRs would be a collection of
scattering centers rather than Kolmogorov turbulence.

However, if future or ongoing experiments in the Northern hemisphere,
e.g. the High Resolution Fly's Eye~\cite{Hires} and AGASA~\cite{AGASA}
keep recording cosmic rays above $\simeq200\,$EeV with large
deflection angles from the line of sight to Cen--A, even the scenario
with $B\simeq1\,\mu$G would be unequivocally ruled out. Similarly, if
no significant anisotropy is seen between these experiments and the
Pierre Auger project~\cite{auger} at the highest energies, the model
would be discarded. As an example, we calculate from Eq.~(\ref{expos})
the fractional difference $\delta I\equiv(I_{\rm N}-I_{\rm S})/(I_{\rm
N}+I_{\rm S})$ of integral fluxes $I_{\rm N}$ and $I_{\rm S}$ that
would be seen by detectors in the Northern and Southern hemispheres
above a given energy (for simplicity, we use exposure functions for
AGASA and an analogous one for the Southern hemisphere, as before).
We find $\delta I\simeq-0.19$ for $E\gtrsim100\,$EeV, and $\delta
I\simeq-0.78$ for $E\gtrsim300\,$EeV. These numbers can also very
roughly been estimated from Fig.~\ref{F4}. Anisotropies of this size
should be easily detectable by a full sky observatory such as the
Pierre Auger project~\cite{auger} which is currently under
construction.

We note in this context that we have assumed in all our
simulations that the source emits only protons. If the source
is sufficiently compact, protons could convert into neutrons
within the source. As pointed out in Ref.~\cite{agw}, for
sources as close as Cen--A, neutrons at the highest energies
could survive decay and produce a spike in the direction
of the source. This can only increase anisotropy. Preliminary
simulations performed with our code indicate that
the total flux in the $\simeq2^\circ$ pixel centered on
the source is significantly (i.e. by a factor $\simeq2$)
increased only above $\simeq300\,$EeV.

We stress that $B\simeq1\,\mu$G corresponds to the upper limit
inferred on the strength of the magnetic field in the Local
Supercluster from Faraday rotation observations of distant
quasars~\cite{vallee,rkb,bbo}. Strictly speaking, the rotation measure
is a function of $B\sqrt{L}$, where $L$ denotes as before the
coherence scale of the field, provided that $L\ll R$, where $R$
represents the size of the medium pervaded by the magnetic field (for
the Local Supercluster $R\simeq10-50\,$Mpc). In principle, the coherence
length $L$ could be smaller than $1\,$Mpc, used in the previous
Section, and $B$ correspondingly larger.  However, as will be
discussed in the next Section, the diffusion coefficient $D(E)$ scales
inversely proportional with $L$, so that by decreasing $L$, one would
increase $D(E)$ correspondingly, and the diffusion approximation would
break down at a smaller energy, thereby increasing the anisotropy at
higher energies. We thus find that one cannot decrease $L$ and
increase $B$ to improve the fit to the data. Moreover, notwithstanding
this fact, if diffusion were to be more efficient at the highest
energies observed, the effective distance traveled would be
consequently increased, and the GZK cut-off would be more pronounced,
which would further aggravate the disagreement with the observed
spectrum.

We thus conclude that $B\simeq1\,\mu$G seems to be the only value of
$B$ that would be marginally consistent with Cen--A as the source
of observed UHECRs. We further note that if
equipartition of energy holds, the magnetic field intensity is tied to
the thermal energy density of the ambient gas:
\begin{equation}
B = 0.5\mu{\rm G}\, T_7^{1/2} \kappa_{10}^{1/2}h_{70},
\end{equation}
where $T_7\equiv T/10^7\,$K is the temperature of the Local
Supercluster in units of $10^7\,$K, and $\kappa_{10}\equiv\kappa/10$,
with $\kappa$ the collapse factor (i.e., the local overdensity of
baryons and electrons), and $h_{70}$ the Hubble constant in units of
$70\,$km/s/Mpc. The gas cools by bremsstrahlung emission in the keV
range, which in principle can be observed. A marginal detection of
X-ray emission correlated with the plane of the Local Supercluster has
actually been reported and corresponds to a collapse factor
$\kappa_{10}\simeq1$, with a weak dependence on the assumed
temperature $T\simeq10^8\,$K~\cite{B99}; the signal is however weak
and these parameters could be in error. Numerical simulations of
large-scale structure formation indicate that $\kappa_{10}\simeq1$ and
$T\simeq10^7\,$K are probably upper limits for sheets such as the
Local Supercluster, and seem to better describe the filamentary
structures~\cite{rkb}. Deep searches for soft X-ray emission correlated
with the plane of the Local Supercluster using the XMM-Newton or
Chandra observatories could improve these limits, and appear
mandatory.

\section{Comparison with analytical estimates}

Let us now compare these results with analytical estimates in an
approach similar to \cite{fp2}. There is often confusion in the
literature about different regimes of diffusion and corresponding
expressions for the diffusion coefficient. It is dangerous to take
analytical expressions too literally as there exists no analytical
derivation of diffusion coefficients in the limit in which the
turbulent component of the magnetic fields becomes comparable or
stronger than a (putative) uniform component, the so-called strong
turbulence regime.  In this respect, one should note that the formula
for the diffusion coefficient given in Ref.~\cite{BO98}, now used
without caution in the community, does not correspond to an analytical
derivation. It is a phenomenological formula, that is furthermore
shown to be in error in Ref.~\cite{CLP01}, where analytical
approximations and accurate measurements of diffusion coefficients
obtained through Monte-Carlo simulations are presented. Notably, this
latter study shows that for Kolmogorov turbulence the diffusion
coefficient can be approximated as:
\begin{eqnarray}
D(E) & \, \simeq \, & 0.02 E_{20}B_{-6}^{-1} \,{\rm Mpc}^2/{\rm
   Myr}\qquad E_{20}< E_{\rm c},\nonumber\\ 
     & \, \simeq \, & 0.02 E_{20}^2B_{-6}^{-2}L^{-1} \,{\rm
   Mpc}^2/{\rm Myr}\qquad E_{20}>E_{\rm c}.
\label{eq_D}
\end{eqnarray}
In this expression, $E_{20}$ is the UHECR energy in units of
$100\,$EeV, $B_{-6}$ is the magnetic field strength in units of
$\mu$G, $L$ is in units of Mpc and $E_{\rm c}=1.45B_{-6}L$ (in units
of $100\,$EeV) corresponds to the condition $r_{\rm L}=L/2\pi$, where
$r_L=0.11E_{20}B_{-6}^{-1}\,$Mpc is the Larmor radius. Note the
difference of the above result with the formula given in
Ref.~\cite{BO98}, for which $D(E)\propto E^{1/3}$ for $E_{20}<E_{\rm
c}$, and $D(E)\propto E$ for $E_{20}>E_{\rm c}$. The dependence of
$D(E)$ for $E<E_{\rm c}$ in Eq.~(\ref{eq_D}) above agrees with the
phenomenological Bohm diffusion coefficient $D_{\rm B}\simeq r_{\rm L}c$
within a factor two.

\begin{figure}
\centering\leavevmode 
\epsfxsize=3.5in
\epsfbox{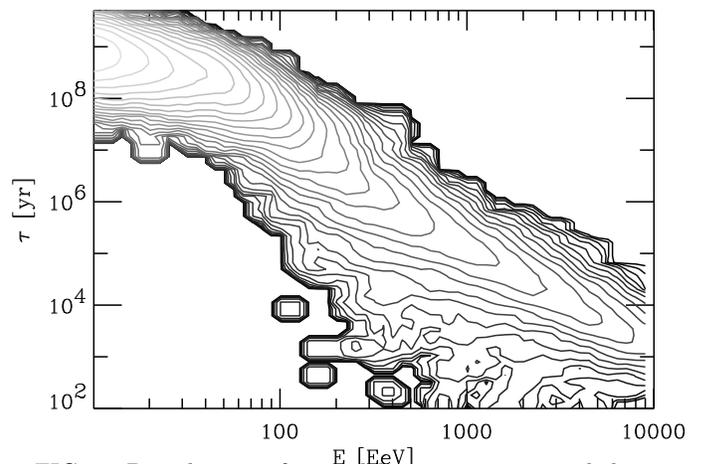} 
\caption[...]{Distribution of time delays $\tau$
versus recorded energy $E$, for $B=1\,\mu$G, and Cen--A as the
source.}
\label{F6}
\end{figure}

The diffusive regime as well as the transition to nearly rectilinear
propagation can also be seen in the dependence of time delay $\tau$
(defined as the difference between the total propagation time and the
straight flight distance $d/c$) on energy $E$ shown in Fig.~\ref{F6}
for $B_{-6}=1$ and Cen--A as the source: In the diffusive regime the
average time delay $\tau(E)\simeq d^2/4D(E)$, where $d$ is the source
distance, whereas for $E_{20}\gtrsim3$, in the regime of almost
rectilinear propagation, $\tau(E)\simeq1.7\times10^{8}\,{\rm
yr}E_{20}^{-2}d_{\rm CenA}^2 LB_{-6}^2$~\cite{WM-E96}, where $d_{\rm
CenA}\equiv d/3.4\,$Mpc.  The values obtained for $D(E)\simeq
d^2/4\tau(E)$ from Fig.~\ref{F6} in the diffusive regime are
consistent with Eq.~(\ref{eq_D}) within the width of the distribution.

Note that according to Eq.~(\ref{eq_D}) a largest eddy size $L$
considerably smaller than 1 Mpc (which is not excluded) would lead
to less diffusion, tending to make anisotropies even larger.

Using these results, one can understand the situation encountered in
the previous section. In the diffusive regime one finds that the ratio
of the diffusive propagation time (which is equal to the time delay in
this approximation) to the source distance reads: $\tau(E)/d\simeq 16
E_{20}^{-2}B_{-6}^2Ld_{\rm CenA}$, for $E_{20}>E_{\rm c}$. Diffusion
ceases to be a good approximation when $\tau(E)/d$ is no
longer $\gg1$. For instance, $\tau(E)/d>2$ requires
$E_{20}<3B_{-6}L^{1/2}$. This corresponds
to a maximal energy for diffusion of $\simeq100\,$EeV when
$B\simeq0.3\,\mu$G, and $\simeq300\,$EeV when $B\simeq1\,\mu$G. These
numbers are in agreement with the results of the previous section.
Note that the ratio of the diffusive propagation time to the source
distance also corresponds to the ratio of the source distance to the
mean free path for scattering on the magnetic inhomogeneities $\simeq
D/c$ within a factor four.

The difference betwen our results and Ref.~\cite{fp2} can be explained
as a misuse of the diffusion coefficient given in Ref.~\cite{BO98} by
the authors of Ref.~\cite{fp2}, and to some extent, by the fact that
that diffusion coefficient itself has the wrong scaling with $E$, $B$,
and $L$. More precisely in Ref.~\cite{fp2} the authors use the
diffusion coefficient $D(E)\simeq 0.05\,{\rm Mpc}^2/{\rm
Myr}E_{20}^{1/3}B_{-6}^{-1/3}L^{2/3}$~\cite{BO98} (note that the
present prefactor is the correct one; their Eq.~(1) has a numerical
error) on the whole energy range, whereas according to
Ref.~\cite{BO98}, this scaling is only valid when $r_{\rm L}\lesssim
L/2\pi$, or as above $E_{20}\lesssim E_c=1.45B_{-6}L$. This makes an
important difference, because, had the authors of Ref.~\cite{fp2} used
the other limiting regime they quote, namely
$D\simeq0.1E_{20}B_{-6}^{-1}\,{\rm Mpc}^2/{\rm Myr}$, valid for
$E_{20}\gtrsim 1.45B_{-6}L$, they would have found that the ratio of
the diffusive distance traveled to source distance reads $d/4D\simeq
2.6E_{20}^{-1}B_{-6}d_{\rm CenA}$, which shows that
$B>1\,\mu$G is necessary to achieve diffusion up to the highest
energies $E_{20}\simeq3$. Instead, they used the former approximation,
giving $d/4D\simeq 5.3E_{20}^{-1/3}B_{-6}^{1/3}L^{-2/3}d_{\rm CenA}$,
which due to the (incorrect) weak scaling would let believe that
diffusion can be achieved easily with $B\simeq 0.3\,\mu$G on the
whole range of energies.

One should mention that the transition between diffusive and
rectilinear regimes of propagation is not sudden; it stretches over
half an order to an order of magnitude. As one approaches this
transition, the anisotropy increases steeply to match the small angle
deflection in the rectilinear regime $\theta_E\simeq140^\circ
E_{20}^{-1}B_{-6}L^{1/2}d_{\rm CenA}^{1/2}$~\cite{WM-E96}. Therefore
it is incorrect to assume that the diffusive estimate for the
anisotropy remains valid up to the transition energy.

Finally, Ref.~\cite{fp2} (unlike Ref.~\cite{agw})
also base their calculations for the flux of
particles detected at Earth on a time-dependent solution of the
diffusion equation, but one should rather use a stationary solution
corresponding to a continuously emitting source. Indeed if Cen--A were
a bursting source, or more generally a source emitting only once on a
timescale $t_{\rm em}\ll\tau_{\rm min}$, where $\tau_{\rm min}$ is the
smallest time delay imparted to all UHECRs, or equivalently, the time
delay at the highest energies observed, an experiment like AGASA would
record events only in a limited range of energies~\cite{M-EW96}. In
other words the distance $d_D$ of the diffusion front from the source
at time $t$ is given by $d_D(E)=6D(E)t$, and depends on energy. At the
present time at which AGASA is operating, the front of low energy
particles would not have yet reached the Earth, while that of higher
energy particles would already have passed. From Fig.~\ref{F6} one
sees that at any given time AGASA would record only part of the total
energy spectrum, since particles with $E=10\,$EeV arrive at time
$\tau\sim10^9\,$yrs, while particles with $E=100\,$EeV arrive at times
$\tau\sim10^8\,$yrs.

This implies that if Cen--A is the source of all UHECRs, it has to be
a continuously emitting source, or, what amounts to the same, an
intermittent source that emits on timescales $t_{\rm em}$, with
quiescent periods of duration $\Delta t \ll {\rm min}[\Delta\tau(E)]$.
Here $\Delta\tau(E)$ denotes the spread of time delays at energy $E$,
and the condition ensures that the various contributions of the
various bursts of particles largely overlap so as to produce a
featureless energy spectrum at all times. Furthermore, since no low
energy cut-off in the energy spectrum has been seen down to energies
$\simeq5\,$EeV, Cen--A must have been active for a time corresponding
to the largest time delay possible $\simeq\tau(5\,{\rm
EeV})+\Delta\tau(5\,{\rm EeV})$.  For $B\simeq1\,\mu$G, we find that
Cen--A must have been producing UHECRs intermittently for the past
$\simeq10\,$Gyrs. Whether this is realistic or not can hardly be
constrained on theoretical grounds. The above constraints on the
duration of the periods of activity and quiescence $t_{\rm em}$ and
$\Delta t$ read $t_{\rm em}\lesssim 10^7\,$yrs, and similarly for
$\Delta t$, which are reasonable orders of magnitude for the evolution
timescale of Cen--A~\cite{I98}. The age of Centaurus~A can be
estimated from the time necessary for its jets to extend to their
present size $\simeq250\,$kpc, i.e. between $\sim10^8\,$yrs and a
${\rm few}\,$Gyrs, depending on the deceleration during extension, so
that for our purposes this age is essentially unknown. Note that
Centaurus~A is a radio-galaxy with sub-relativistic jets, and without
hot spots; the lobes are not particularly active, with a total
bolometric luminosity $\sim 10^{39}\,$ergs/s~\cite{I98}.

The constraints of Ref.~\cite{fp2} on the energy requirement must thus
be reconsidered. In the stationary regime, the flux at Earth reads
$E^2j(E)=E^2q(E)/(4\pi)^2D(E)d$, where $q(E)$ is the injection
spectrum at the source. One easily calculates that, assuming
$q(E)\propto E^{-2}$, in order to produce the energy weighted flux
measured by cosmic ray experiments $\simeq
10^{24.5}\,$eV$^2$m$^{-2}$s$^{-1}$sr$^{-1}$, one requires a UHECR
emission power ${\cal P}_{\rm UHECR}\simeq 10^{39}B_{-6}^{-1}d_{\rm
CenA}\,$ergs/s at the source. Note that this represents the average
power; the actual power during the phase of activity is higher and
reads ${\cal P}_{\rm UHECR}(t_{\rm em}+\Delta T)/t_{\rm em}$.  The
above ${\cal P}_{\rm UHECR}$ is thus a strict lower limit, which is
nevertheless more optimistic than the $\sim10^{43}\,$ergs/s obtained
in Ref~\cite{fp2}.

As an aside, we note that the above constraint on the age of
Centaurus~A can be generalized to any such source of UHECRs, namely
that the time delay at the lowest energies be smaller than the age of
the source, and thus also the age of the Universe. The propagation
time at $5\,$EeV reads, using Eq.~(\ref{eq_D}): $\tau\simeq2.8\,{\rm
Gyrs}\,B_{-6}d_{\rm CenA}^2$. Imposing $\tau\leq 14\,$Gyrs gives a
general constraint between the distance $d$ to the source and the
strength of the intervening magnetic field:
\begin{equation}
\left(\frac{B}{1\,\mu{\rm G}}\right)
\left(\frac{d}{10\,{\rm Mpc}}\right)^2\lesssim0.6
\end{equation}

Even if there are several sources contributing to the cosmic ray flux,
this limit should hold unless the sources conspire to add their
individual piecewise contribution in such a way as to form a
featureless energy spectrum. When there are many sources the above
constraint disappears, as the central limit theorem would guarantee
that a featureless spectrum would be produced; this is notably one of
the peculiarities of the $\gamma-$ray burst model of UHECR
origin~\cite{waxman}.

\section{Conclusions}

Our detailed numerical simulations show that the model considered in
Ref.~\cite{fp2}, in which Centaurus~A is the source of all observed
UHECRs, is inconsistent with the data, at least for the magnetic field
strength $B\simeq0.3\,\mu$G put forward by these authors. We find that
for a magnetic field strength $B\simeq1\,\mu$G, the predicted energy
spectrum is in marginal agreement with that observed by AGASA. However
the large deflection angle of the highest energy event (the Fly's Eye
event) with respect to the line of sight to Cen--A must be explained
as a $\simeq2\sigma$ fluctuation. We also argued that this magnetic
field intensity saturates the observational upper bounds from Faraday
rotations and on X-ray emission from the ambien gas. This model can be
tested by improving these upper limits with current experiments. We
also showed that in order to explain all UHECRs down to
$E\simeq5\,$EeV, Cen--A must have been producing UHECRs for the past
$\simeq10\,$Gyrs. All these facts are rather strong requirements on
the source and on the intervening magnetic fields. We thus find anew a
conclusion obtained in previous studies~\cite{ES95,BO98,SLB99,LSB99},
namely it is much more likely that a few sources within $\simeq10\,$Mpc from
the Earth would produce the observed ultra-high energy cosmic ray
flux, and that the ambient magnetic field strength in the Local
Supercluster $B\simeq0.1-0.5\,\mu$G. Work is in progress to quantify
the number of sources needed and the distance scale for various values
of the magnetic field strength. Ongoing and future ultra-high energy
cosmic ray experiments~\cite{Hires,AGASA,auger}, by increasing the
statistics at the highest energies, will soon provide much tighter
bounds on the number of UHECR sources.

\section*{Acknowledgements}
G.S. would like to thank Luis Anchordoqui, Pasquale Blasi,
Glennys Farrar, Haim Goldberg, Tsvi Piran, and Tom Weiler
for detailed discussions of this subject.


\end{multicols}


\begin{thebibliography}{9}  

\bibitem{gzk} K.~Greisen, Phys.~Rev.~Lett. 16 (1966)
748; G.~T.~Zatsepin and V.~A.~Kuzmin, Pis'ma
Zh. Eksp. Teor. Fiz. 4 (1966) 114 [JETP. Lett. 4 (1966) 78].

\bibitem{Fly} D. J. Bird et al., Phys. Rev. Lett.71 (1993) 3401~;
  Astrophys. J. 424 (1994) 491~; ibid.441 (1995) 144.

\bibitem{Haverah} See, e.g., M. A. Lawrence, R. J. O. Reid, and
  A. A. Watson, J.Phys. G 17 (1991) 733, and references
  therein~; see also http~://ast.leeds.ac.uk/haverah/hav-home.html

\bibitem{Yakutsk} N. N. Efimov et al., Proc. International Symposium on
  {\it Astrophysical Aspects of the Most Energetic Cosmic
  Rays}, eds.  M. Nagano and F. Takahara (Worls Scientific Singapore,
  1991) p.20~; B. N. Afnasiev, Proc. of International Symposium on {\it
  Extremely High Energy Cosmic Rays~: Astrophysics and Future
  Observatoires}, ed. M. Nagano (Instiute for Cosmic Ray Research,
  Tokyo, 1996), p.32.

\bibitem{Hires} D. Kieda et al., Proc. of the 26th ICRC, Salt Lake,
  1999~; www.physics.utah.edu/Resrch.html

\bibitem{AGASA} Takeda et al., Astrophys. J. 522 (1999) 225;
M.Takeda et al., Phys. Rev. Lett. 81 (1998) 1163; Hayashida et al.,
e-print astro-ph/0008102; www-akeno.icrr.u-tokyo.ac.jp/AGASA/.

\bibitem{reviews} for recent reviews see J.~W.~Cronin, Rev.~Mod.~Phys.
71 (1999) S165; P.~Bhattacharjee and G.~Sigl, Phys.~Rept. 327 (2000) 109;
A.~V.~Olinto, Phys.~Rept. 333-334 (2000) 329; X.~Bertou,
M.~Boratav, and A.~Letessier-Selvon, Int.~J.~Mod.~Phys. A15 (2000) 2181;

\bibitem{biermann} see, e.g., P.~L.~Biermann, {\it J.~Phys.~G:
Nucl.~Part.~Phys.} {\bf 23}, 1 (1997).

\bibitem{ssb} G.~Sigl, D.~N.~Schramm, and P.~Bhattacharjee,
Astropart.~Phys. 2 (1994) 401

\bibitem{ES95} J.~W.~Elbert, and P.~Sommers, Astrophys.~J. 441
(1995) 151;

\bibitem{WM-E96} E.~Waxman and J.~Miralda-Escud\'{e},
Astrophys.~J. 472 (1996) L89.

\bibitem{vallee} J.~P.~Vall\'{e}e, Fundamentals of Cosmic
Physics, Vol.~19 (1997) 1.

\bibitem{rkb} D.~Ryu, H.~Kang, and P.~L.~Biermann,
Astron.~Astrophys. 335 (1998) 19.

\bibitem{bbo} P.~Blasi, S.~Burles, and A.~V.~Olinto, Astrophys.~J.
514 (1999) L79.

\bibitem{BO98} P.~Blasi and A.~V.~Olinto, Phys.~Rev.~D. 59 (1999)
023001.

\bibitem{SLB99} G.~Sigl, M.~Lemoine, and P.~Biermann, Astropart.~Phys.
10 (1999) 141.

\bibitem{LSB99} M.~Lemoine, G.~Sigl, P.~Biermann, e-print astro-ph/9903124.

\bibitem{ambs} E.-J.~Ahn, G.~Medina-Tanco, P.~L.~Biermann,
and T.~Stanev, e-print astro-ph/9911123.

\bibitem{fp2} G.~R.~Farrar and T.~Piran, e-print astro-ph/0010370.

\bibitem{M-TE00} G.~Medina-Tanco and T.~A.~Ensslin, Astropart.~Phys.,
to appear, e-print astro-ph/0011454.

\bibitem{auger} J.~W.~Cronin, Nucl.~Phys.~B (Proc.~Suppl.) 28B
(1992) 213; The Pierre Auger Observatory Design Report (2nd
edition), March 1997; see also
{\sf http://http://www.auger.org/} and
{\sf http://www-lpnhep.in2p3.fr/auger/welcome.html}.

\bibitem{agw} L.~A.~Anchordoqui, H.~Goldberg, and T.~Weiler,
e-print astro-ph/0103043.

\bibitem{B99} S.P.~Boughn, ApJ 526, 14 (1999).

\bibitem{CLP01} F.~Cass\'e, M.~Lemoine and G.~Pelletier, in preparation.

\bibitem{M-EW96} J.~Miralda-Escud\'{e} and E.~Waxman, Astrophys.~J.
462 (1996) L59.

\bibitem{I98} F.~P.~Israel, Astron.~Astrophys.~Rev. 8, 237 (1998).

\bibitem{waxman} E.~Waxman, {\it Phys.~Scripta} {\bf T85}, 117 (2000),
and references therein.


\end{thebibliography}
\end{document}